\documentstyle[prb,aps,preprint,epsf,amsfonts,amssymb]{revtex}

\begin{document}


\title{Disorder-induced phonon self-energy of semiconductors\\ with binary
isotopic composition}
\author{F.~Widulle, J.~Serrano, and M.~Cardona}
\address{Max-Planck-Institut~f\"{u}r~Festk\"{o}rperforschung, Heisenbergstr.~1,
D-70569 Stuttgart, Germany}
\date{May 29, 2001}

\maketitle

\begin{abstract}
\baselineskip=21pt
Self-energy effects of Raman phonons in isotopically disordered
semiconductors are deduced by perturbation theory and compared to
experimental data. In contrast to the acoustic frequency region, higher-order
terms contribute significantly to the self-energy at optical phonon
frequencies. The asymmetric dependence of the self-energy of a binary isotope
system $\rm{m}_{1-x}\rm{M}_x$ on the concentration of the heavier isotope
mass~$x$ can be explained by taking into account second- and third-order
perturbation terms. For elemental semiconductors, the maximum of the
self-energy occurs at concentrations with $0.5<x<0.7$, depending on the
strength of the third-order term. Reasonable approximations are imposed that
allow us to derive explicit expressions for the ratio of successive
perturbation terms of the real and the imaginary part of the self-energy. This
basic theoretical approach is compatible with Raman spectroscopic results on
diamond and silicon, with calculations based on the coherent potential
approximation, and with theoretical results obtained using {\it ab initio}
electronic theory. The extension of the formalism to binary compounds, by
taking into account the eigenvectors at the individual sublattices, is
straightforward. In this manner, we interpret recent experimental results on
the disorder-induced broadening of the TO (folded) modes of SiC with a
$^{13}{\rm C}$-enriched carbon sublattice.~\cite{Rohmfeld00,Rohmfeld01}
\end{abstract}

\pacs{PACS numbers: 78.30.-j, 63.20.-e, 63.50.+x, 78.30.Ly}

\section{Introduction}
\label{sec:intro}

The mass-fluctations (of a single element) induced by isotopic disorder can be
described by the weighted $n$-th moments of the relative mass
differences with respect to the average mass $\overline{m}=\sum_i c_i m_i$ of
the virtual crystal defined as

\begin{equation}
\label{gn-general}
g_n=
\sum\limits_i\,
c_i\,
\left(
\frac{\overline{m}-m_i}{\overline{m}}
\right)^n,
\end{equation}

where $i$ runs over the different isotopes with masses $m_i$ and
concentrations $c_i$. According to Eq.~(\ref{gn-general}), $g_1=0$ within the
virtual crystal approximation (VCA), i.e. if $\overline{m}$ is used as the
atomic mass.

For a two-isotope system $m_{1-x}M_x$ with $M$ being the heavy mass and
$x=c_M$, Eq.~(\ref{gn-general}) reads

\begin{equation}
\label{gn-binary}
g_n(x)=
(1-x)\,
\left(
\frac{\overline m-m}{\overline{m}}
\right)^n+
x\,
\left(
\frac{\overline m-M}{\overline{m}}
\right)^n.
\end{equation}
Equation~(\ref{gn-binary}) can be rewritten as follows, using $\Delta
m=M-m$

\begin{eqnarray}
\label{gn-binary2}
g_n(x) &=&
\left(m/\Delta m+x\right)^{-n}\nonumber\\
 & & \times x(1-x)\left(x^{n-1}+(-1)^n (1-x)^{n-1}\right).
\end{eqnarray}

For the elements under consideration here, the first factor of
Eq.~(\ref{gn-binary2}) leads to a weak $x$-dependence that slightly
breaks the odd or even symmetry of $g_n(x)$ with respect to $x=0.5$.
Neglecting this dependence (i.e. for $x\ll m/\Delta m$) we can write

\begin{equation}
\label{gn-factor}
\left(m/\Delta m+x\right)^{-n}\simeq (\Delta m/m)^n.
\end{equation}

The lowest-order moments of $g_n$ thus read (see Fig.~\ref{fig:gn}):
\begin{mathletters}
\label{g2345}
\begin{eqnarray}
\label{g2}
g_2 &\simeq & (\Delta m/m)^2\,x\,(1-x)\\
\label{g3}
g_3 &\simeq & (\Delta m/m)^3\,x\,(1-x)\,(2x-1)\\
\label{g4}
g_4 &\simeq & (\Delta m/m)^4\,x\,(1-x)\,(1-3x(1-x))\\
\label{g5}
g_5 &\simeq & (\Delta m/m)^5\,x\,(1-x)\,(2x-1)\,(1-2x(1-x))
\end{eqnarray}
\end{mathletters}

Hence, the contribution to the self-energy is even (odd) with respect to
$x=0.5$ for even (odd) values of $n$. In particular, $g_2$ represents a
parabola with the maximum at $x=0.5$, while $g_3$ is odd with respect to
$x=0.5$, i.e. $g_3(0.5)=0$.

The disorder-induced phonon self-energy is defined as
\begin{equation}
\label{self-energy}
\Pi_{\rm{dis}}(\omega)=\Delta_{\rm{dis}}(\omega)-i\,\Gamma_{\rm{dis}}(\omega)/2,
\end{equation}
where $\Delta_{\rm{dis}}$ denotes the disorder-induced shift, and
$\Gamma_{\rm{dis}}$ the disorder-induced broadening corresponding to the
full width at half maximum (FWHM). The self-energy can be decomposed into
individual contributions that are proportional to the various moments of $g_n$
with $n=2,3,4,\ldots$

\begin{eqnarray}
\label{gn-fit}
\Pi_{\rm{dis}}(\omega)
&=& ~~\,A_2\,x\,(1-x)\nonumber\\
& & +A_3\,x\,(1-x)\,(2x-1)\nonumber\\
& & +A_4\,x\,(1-x)\,(1-3x(1-x))\nonumber\\
& & +\ldots
\end{eqnarray}

In order to estimate the contributions of higher-order perturbation terms,
Eq.~(\ref{gn-fit}) is fitted to experimental and theoretical data of
group-IV semiconductors (see Sec.~\ref{sec:comparison} for diamond, Si, Ge,
$\alpha$-Sn and Sec.~\ref{sec:compounds} for SiC). In Eq.~(\ref{gn-fit}) the
fit parameters $A_i=A_{i,\Delta}+i A_{i,\Gamma}$ are complex numbers; in
practice, we perform individual fits to $\Delta_{\rm dis}$ and $\Gamma_{\rm
dis}$, thus obtaining the two sets of fit parameters $A_{i,\Delta}$ and
$A_{i,\Gamma}$. In the following section we show that, in general, the
asymmetries with respect to $x=0.5$ are not identical for $\Delta_{\rm{dis}}$
and $\Gamma_{\rm{dis}}$.

For a characterization of the contribution of the third-order perturbation
term with respect to the second-order term, we define the ratio

\begin{equation}
\label{ratio}
r_s=\frac{A_{3,s}}{A_{2,s}}
\end{equation}
with $s=\Delta,\Gamma$. For the maximum self-energy $x_{{\rm max},s}$ we
find

\begin{equation}
\label{xmax}
x_{{\rm max},s}=
\frac{1}{2}\pm\frac{1}{6}\left(\sqrt{3+\frac{1}{r_s^2}}-\frac{1}{|r_s|}\right),
\end{equation}
where the positive (negative) sign corresponds to $r>0$ ($r<0$).

\section{Perturbation theory}
\label{sec:theory}

In this section, explicit expressions for the disorder-induced self-energies of
a two-isotope elemental crystal are derived by perturbation theory. For the
calculations we use $\Pi_{\rm{dis}}=\Pi=\Delta+i \Gamma$ with
$\Delta=\Delta_2+\Delta_3+\ldots=\Delta_{\rm{dis}}$ and
$\Gamma=\Gamma_2+\Gamma_3+\ldots=-\Gamma_{\rm{dis}}/2$. The zero-temperature
phonon propagator~\cite{Mahan} $D(\omega,\omega_i)$ for an isotopically pure
crystal is the sum of the propagators $D_+(\omega,\omega_i)$ and
$D_-(\omega,\omega_i)$ for positive and negative frequencies, respectively,

\begin{mathletters}
\begin{eqnarray}
D &=& D_+(\omega,\omega_i)+D_-(\omega,\omega_i)\\
&=& \frac{1}{\omega-\omega_i-i\gamma}+\frac{1}{-\omega-\omega_i-i\gamma}\\
&=& \frac{2\omega_i}{\omega^2-\omega_i^2-i2\omega_i\gamma},
\end{eqnarray}
\end{mathletters}

where $\gamma\rightarrow 0$ if anharmonicity is neglected. $\omega_i$
denotes the absolute value of the phonon frequency. Since we are interested in
the self-energy of the Raman mode, to which the largest contributions stem from
close-lying optic phonon states, the propagator $D_-$ can be neglected because
$|{\rm Re}\{D_-\}|\ll |{\rm Re}\{D_+\}|$ and $|{\rm Im}\{D_-\}|\ll |{\rm
Im}\{D_+\}|$. In the following we thus set $D=D_+$ with

\begin{mathletters}
\begin{eqnarray}
D_+(\omega,\omega_i)
&=& \frac{1}{\omega-\omega_i-i\gamma}\\ 
&=& \frac{\omega-\omega_i}{(\omega-\omega_i)^2+\gamma^2}
+i\,\frac{\gamma}{(\omega-\omega_i)^2+\gamma^2}\\  
&=& P_i(\omega)+i L_i(\omega).
\end{eqnarray}
\end{mathletters}

$D_+(\omega,\omega_i)$ consists of $P_i(\omega)$,
with a pole at $\omega_i$, and a Lorentzian $L_i(\omega)$. For the complex
self-energy of the Raman phonon of a semiconductor with diamond structure,
$n$th-order perturbation theory yields~\cite{Tamura83}

\begin{mathletters}
\begin{eqnarray}
\Pi_n(\omega,x)
&=&
\frac{g_n(x)}{2^n}\,\omega
\left(\frac{1}{6N_c}\right)^{n-1}\nonumber\\
& & \times\sum\limits_{i,j,...,n}
\omega_i D_i\cdot
\omega_j D_j\cdots
\omega_n D_n\\
&=&
\frac{g_n(x)}{2^n}\,\omega \left(\frac{1}{6N_c}\sum\limits_i
\omega_i D_i\right)^{n-1}
\end{eqnarray}
\end{mathletters}

with $D_n=D(\omega,\omega_n)$ and $N_c$ the number of unit cells.
The second-order term ($n=2$)

\begin{mathletters}
\begin{eqnarray}
\Pi_2(\omega,x)
&=&
\frac{g_2(x)}{4}\,\omega
\left(
\frac{1}{6N_c}\sum\limits_i \omega_i (P_i+i L_i)
\right)
\end{eqnarray}
contains the real part

\begin{equation}
\label{delta_2}
\Delta_2(\omega,x)=
\frac{g_2(x)}{4}\,\omega
\left(
\frac{1}{6N_c}\sum\limits_i
\omega_i\frac{\omega-\omega_i}{(\omega-\omega_i)^2+\gamma^2}
\right)
\end{equation}
and the imaginary part

\begin{equation}
\label{gamma_2}
\Gamma_2(\omega,x)
=
\frac{g_2(x)}{4}\,\omega
\left(
\frac{1}{6N_c}\sum\limits_i
\omega_i \frac{\gamma}{(\omega-\omega_i)^2+\gamma^2}
\right).
\end{equation}

For $\gamma\rightarrow 0$, Eq.~(\ref{gamma_2}) simplifies to a sum over
$\delta$-functions, which represents the one-phonon density of states (DOS)
$\rho_1(\omega)$,

\begin{eqnarray}
-\Gamma_2(\omega,x) &=&
\frac{g_2(x)}{4}\,\omega^2
\left(
\frac{1}{6N_c}\sum\limits_i  \pi\,\delta(\omega-\omega_i)
\right)\\
\label{gamma_2-formula}
&=& \frac{\pi}{24}\,g_2(x)\,\omega^2\,\rho_1(\omega).
\end{eqnarray}
\end{mathletters}

If $\Delta_2(\omega)$ and $\Gamma_2(\omega)$ do not depend strongly on
$\omega$ near the Raman phonon at $\omega_0$, then $-\Gamma_2(\omega_0)$ of
Eq.~(\ref{gamma_2-formula}) represents the half width at half maximum (HWHM) of
the disorder-induced broadening.

The third-order self-energy can be written as

\begin{mathletters}
\begin{eqnarray}
\Pi_3(\omega,x)
&=&
\frac{g_3(x)}{8}\,\omega
\left(\frac{1}{6N_c}\right)^2\nonumber\\
& &
\times\sum\limits_{i,j} \omega_i \omega_j (P_i + i L_i)(P_j +i L_j),
\end{eqnarray}
yielding the following real and imaginary parts

\begin{eqnarray}
\Delta_3(\omega,x)
&=&
\frac{g_3(x)}{8}\,\omega
\left(\frac{1}{6N_c}\right)^2\nonumber\\
& & \times\sum\limits_{i,j} \omega_i \omega_j (P_i P_j-L_i L_j),\\
\Gamma_3(\omega,x)
&=&
\frac{g_3(x)}{4}\,\omega
\left(\frac{1}{6N_c}\right)^2
\sum\limits_{i,j} \omega_i \omega_j P_i L_j.
\end{eqnarray}
\end{mathletters}

The corresponding fourth-order terms read

\begin{mathletters}
\begin{eqnarray}
\Delta_4(\omega,x) &=&
\frac{g_4(x)}{16}\,\omega
\left(\frac{1}{6N_c}\right)^3\nonumber\\
& & \times\sum\limits_{i,j,k} \omega_i \omega_j \omega_k (P_i P_j P_k-3 L_i L_j
P_k),\\
\Gamma_4(\omega,x) &=&
\frac{g_4(x)}{16}\,\omega
\left(\frac{1}{6N_c}\right)^3\nonumber\\
& & \times\sum\limits_{i,j,k} \omega_i \omega_j \omega_k (3 P_i P_j L_k-L_i L_j
L_k).
\end{eqnarray}
\end{mathletters}

The ratios of the third-order to the second-order perturbation terms for the
real and imaginary part of the self-energy is simply related to the
disorder-induced shift $\Delta_2$ and broadening $\Gamma_2$:

\begin{mathletters}
\begin{eqnarray}
\frac{\Delta_3}{\Delta_2}
&=&
~\,\frac{~g_3(x)}{2 g_2(x)}
~\frac{1}{6N_c}
\frac{\left(\sum\limits_i \omega_i P_i\right)^2- \left(\sum\limits_i
\omega_i L_i\right)^2}{\sum\limits_i \omega_i P_i}\nonumber \\
\label{r3delta}
&=& 2\,\frac{g_3(x)}{g_2(x)}
\,\frac{1}{\omega}
\left[\frac{\Delta_2^2-\Gamma_2^2}{g_2 \Delta_2}\right]_{x=x_0}\\
\frac{\Gamma_3}{\Gamma_2}
&=&
~\,\frac{g_3(x)}{g_2(x)}
~\frac{1}{6N_c}
\sum\limits_i \omega_i P_i\nonumber \\
\label{r3gamma}
&=& 4\,\frac{g_3(x)}{g_2(x)}
\,\frac{1}{\omega}
\left[\frac{\Delta_2}{g_2}\right]_{x=x_0}
\end{eqnarray}
\end{mathletters}

The ratios $r_s$ (Eq.~(\ref{ratio})) for a certain phonon mode at $\omega_0$
can be determined from the dilute limits $x\rightarrow \{0,1\}$ and the
experimental shifts and broadenings for $x_0=0.5$. Under the approximation of
Eq.~(\ref{gn-factor}), the expressions of Eq.~(\ref{r3delta}) and
Eq.~(\ref{r3gamma}) simplify to

\begin{mathletters}
\begin{eqnarray}
\label{r3delta-final}
r_{\Delta}
&\simeq &
~\,8\,\frac{m}{\Delta m}\,\frac{1}{\omega}
\left[\frac{\Delta_2^2-\Gamma_2^2}{\Delta_2}\right]_{x_0=0.5}\\
\label{r3gamma-final}
r_{\Gamma}
&\simeq &
16\,\frac{m}{\Delta m}\,\frac{1}{\omega}\left[\Delta_2\right]_{x_0=0.5}
\end{eqnarray}
\end{mathletters}

For phonons of crystals whose $|\Delta_2(\omega_0,x_0)|$ and
$|\Gamma_2(\omega_0,x_0)|$ are comparable, such as for the Raman mode of
diamond or the TO Raman modes of SiC, the asymmetry of $\Delta(\omega_0,x)$
also depends on $\Gamma_2(\omega_0,x_0)$ and $r_{\Gamma}/r_{\Delta}\propto
2\Delta_2^2/(\Delta_2^2-\Gamma_2^2)$ ($r_{\Gamma}/r_{\Delta}\approx 8/3$ for
the Raman mode of diamond).

For phonons with $|\Gamma_2(\omega_0)|\ll|\Delta_2(\omega_0)|$, such as the
Raman modes of Si, Ge, and $\alpha$-Sn, Eq.~(\ref{r3delta-final}) can be
further simplified. In these cases, $r_{\Delta}\propto\Delta_2$ and
$r_{\Gamma}/r_{\Delta}\simeq 2$. The fact that, generally, the contributions
of higher-order terms to the self-energy are not equally distributed among the
real and the imaginary part ($r_{\Delta}\ne r_{\Gamma}$) means, that the
asymmetries of $\Delta(\omega_0,x)$ and $\Gamma(\omega_0,x)$ have different
shapes resulting in different concentrations of maximum self-energies
($x_{\rm{max},\Delta}\ne x_{\rm{max},\Gamma}$). The ratios of the next
successive-order terms read

\begin{equation}
\frac{A_{4,\Gamma}}{A_{3,\Gamma}}= \frac{3}{4}\,
\frac{A_{3,\Gamma}}{A_{2,\Gamma}}= \frac{3}{2}\,
\frac{A_{4,\Delta}}{A_{3,\Delta}}= \frac{3}{2}\,
\frac{A_{3,\Delta}}{A_{2,\Delta}}.  
\end{equation}

Figure~\ref{fig:xmax} shows the dependence on $r_s$ of the concentration
$x_{\rm max}$, at which the self-energy has a maximum, if
$|\Gamma_2(\omega_0)|\ll|\Delta_2(\omega_0)|$ (valid for Si,
Ge, and $\alpha$-Sn).

General expressions for the ratios of successive-order perturbation terms for
phonons with $|\Gamma_2(\omega_0)|\ll|\Delta_2(\omega_0)|$ can be easily found
using

\begin{eqnarray}
\left(\sum\limits_i \omega_i\,D_i\right)^n &\propto &
(P+i L)^n=\eta^n e^{i n \theta}\nonumber\\
&=&\eta^n (\cos(n\theta)+i \sin(n\theta))\simeq
\eta^n (1+i n\theta),
\end{eqnarray}
where $\eta=|D|$ and $\theta=\rm{Arg}\{D\}$. They are:

\begin{mathletters}
\begin{eqnarray}
\frac{\Delta_{n+1}}{\Delta_n} &=& \frac{g_{n+1}}{2 g_n}
\left(\frac{1}{6N_c} \sum\limits_i \omega_i P_i \right)\\
\frac{\Gamma_{n+1}}{\Gamma_n} &=& \frac{g_{n+1}}{2 g_n}
\frac{n}{n-1} \left(\frac{1}{6N_c} \sum\limits_i \omega_i P_i \right)\\
\label{rngamma}
&=& \frac{n}{n-1}\frac{\Delta_{n+1}}{\Delta_n}
= \frac{n}{n-1}\frac{\Delta_{3}}{\Delta_2}
\end{eqnarray}
\end{mathletters}
According to Eq.~(\ref{r3delta-final}), negative values for $r_{\Delta}$ are
possible if $|\Gamma_2|$ exceeds $|\Delta_2|$, but this does not occur for
the Raman phonons of elemental semiconductors. One could also obtain
$r_{\Gamma}<0$ (Eq.~(\ref{r3gamma-final})) for phonons with ${\bf q}\ne 0$
that exhibit a negative disorder-induced shift.\cite{note1} The latter
originates from the fact that the shift is related to the Kramers-Kronig
transformation of the broadening, which displays a change of sign of
$\Delta_{\rm dis}(\omega)$ near maxima of $\Gamma_{\rm dis}(\omega)$. In such
a case the asymmetry would be flipped with respect to $x=0.5$, i.e. the
maximum of the self-energy would occur at a concentration $x_{\rm{max}}<0.5$.

\section{Comparison with experimental and theoretical results}
\label{sec:comparison}

In this section, we display a compilation of the disorder-induced
self-energies for the Raman phonons of elemental crystals (diamond, Si, Ge,
$\alpha$-Sn) which have been obtained either by Raman spectroscopy or from
theoretical calculations by several research groups during the past decade.
Raman studies that address the variation of the self-energy with the
isotopic composition have been conducted for
diamond~\cite{Chrenko88,Hass91,Hass92,Spitzer93,Hanzawa96,Vogelgesang96} and
Si.\cite{Widulle01} The coherent potential approximation (CPA) has been
employed for diamond~\cite{Hass91,Hass92,Spitzer93} and Si~\cite{Widulle01},
while {\it ab initio} electronic structure based calculations have been
performed for diamond and Ge.\cite{Vast00a,Vast00b} All sets of data are
analyzed with respect to the asymmetries of the self-energies and compared to
the results we obtain by perturbation theory.

Figure~\ref{fig:DiamondShift} displays the Raman frequencies of diamond versus
the $^{13}{\rm C}$-concentration. The points (open symbols) represent
experimental values. The dashed curve represents the approximately linear
dependence expected in the VCA. The upward curvature of the experimental data
(with respect to the VCA line) clearly demonstrates the existence of an
isotopic-disorder-induced self-energy as emphasized by the solid line, which
is a fit with Eq.~(\ref{gn-fit}) for $n=2, 3$. The prefactors $A_2$ and $A_3$
contain all factors independent of $x$. The ratios $r_s=A_3/A_2$ of the fitted
values of these prefactors are listed in Table~\ref{tab:comparison} under
``Raman experiments''. Note that two points, which are represented by dashed
triangles, have not been included in the fit since they deviate unreasonably
from the other data.

It is difficult to see with the naked eye in Fig.~\ref{fig:DiamondShift} the
asymmetric behavior versus $x$, which may arise from third-order perturbation
terms. The asymmetry appears, however, rather clearly when the difference
between the measured (or the calculated) behavior and the VCA line is plotted,
as shown in Fig.~\ref{fig:DiamondDelta}. In this figure, the solid line also
represents the fit to all experimental data (except the dotted triangles), the
dot-dashed line represents CPA calculations while the dotted line is a fit to
the asterisks which indicate points obtained in the {\it ab initio}
calculations.\cite{Vast00a,Vast00b} All data in Fig.~\ref{fig:DiamondDelta}
show a similar asymmetric behavior, with a maximum of $\Delta_{\rm dis}(x)$ at
$x\approx 0.6$. This figure allows us to conclude that the real part of the
self-energy due to isotopic disorder is well understood for diamond, including
the superposition of second-order and third-order perturbation terms.

Similar degree of understanding has been reached for $\Gamma_{\rm dis}(x)$ as
shown in Fig.~\ref{fig:DiamondGamma}. The $x$-position of the maxima of
$\Delta_{\rm dis}$ and $\Gamma_{\rm dis}$ determined from the experimental data
agree with those obtained by perturbation theory (Eq.~(\ref{xmax})) and also
with the CPA and {\it ab initio} calculations (see Table~\ref{tab:comparison}).

Concerning the other elemental semiconductors, detailed experimental results
with sufficient values of $x$ to reach quantitative conclusions of the type
found for diamond, are only available for Si.\cite{Widulle01} These data for
Si are shown in Fig.~\ref{fig:Si_Delta} and Fig.~\ref{fig:Si_Gamma} (filled
circles) together with the results of CPA calculations (filled squares). The
latter show for $\Delta_{\rm dis}(x)$ a clear asymmetry with a maximum at
$x_{\rm{max},\Delta}\approx 0.56$. The quality and the number of the
experimental points are not sufficient to conclude that an asymmetry exists
but they cannot exclude it either. The measured absolute values of
$\Delta_{\rm dis}(x)$ almost (not quite) agree within error bars with the
calculated ones.

The corresponding experimental values of $\Gamma_{\rm dis}(x)$ (see
Fig.~\ref{fig:Si_Gamma}) are about a factor of two lower than the calculated
ones, although both show the asymmetric behavior ($x_{\rm{max},\Gamma}\approx
0.62$) predicted by Eq.~(\ref{r3gamma-final}). The reason for the discrepancy
between the calculated and measured $\Gamma_{\rm dis}$ is to be sought in the
mechanism responsible for it in Si.\cite{Widulle01} Within the harmonic
approximation $\Gamma_{\rm dis}=0$ for Si, Ge, and $\alpha$-Sn, because the
Raman frequency is at the maximum of the spectrum and thus corresponds to zero
density of one-phonon states. The rather small, but non-negligible, observed
value of $\Gamma_{\rm dis}$ results from the DOS induced at the
$\Gamma$-point by the anharmonic interactions responsible for the linewidth
of the isotopically pure crystals. Thus, the widths observed for Si, as well
as for Ge and $\alpha$-Sn, correspond to fourth-order (twice disorder and
twice anharmonicity) and higher-order terms. Under these conditions, we have
no guarantee that the terms included in an anharmonic CPA calculation suffice
to describe the experimental data, and agreement within a factor of two
between theory and the small experimental values of $\Gamma_{\rm dis}$ is to
be regarded as satisfactory at this point. Using this argument too, similar
values of $\Delta_{\rm{dis}}$ are to be expected for these elemental crystals.

Although isotopic-disorder-induced effects have not been studied so
extensively in Ge and $\alpha$-Sn, there are some experimental results
available for $x=0.5$~\cite{Zhang98,Wang97} that can be compared with
perturbation calculations. The latter, once contrasted with the measurements
for $x=0.5$, provide reliable predictions for future experiments. The
corresponding data are displayed in Table~\ref{tab:comparison}, where we
compared them, in the case of Ge, with those obtained by {\it ab initio}
calculations.\cite{Vast00a,Vast00b} We should add that the {\it ab initio}
values of $\Delta_{\rm dis}$ for Ge calculated by Vast and
Baroni~\cite{Vast00a,Vast00b} agree with the CPA calculation on Si and also
represent the available experimental data rather well
($\Delta^{ab\,initio}(x=0.5)=1.14~\rm{cm}^{-1}$ compared to
$1.18~\rm{cm}^{-1}$ and $1.06~\rm{cm}^{-1}$ obtained experimentally for
$\Delta_2$ in Si and Ge, respectively). The {\it ab initio} calculations of
$\Gamma_{\rm dis}$ for Ge, however, give values which are an order of
magnitude larger than the experimental ones, a fact that can be attributed to
numerical errors in the calculated DOS near the van Hove singularity which is
found at the Raman frequency.

According to the values listed in Table~\ref{tab:comparison} one can clearly
distinguish between two different behaviors in the magnitudes of $\Delta_2$ and
$\Gamma_2$. Diamond exhibits a relatively large disorder-induced shift,
$\Delta_2$, whereas the other elemental crystals have much smaller values.
This largely reflects the factor of $\omega$ in Eq.~(\ref{delta_2}).
The difference is even more striking for $\Gamma_2$, which nearly vanishes for
Si, Ge and $\alpha$-Sn. This can be explained also by the different orders of
perturbation theory coming into play.  The overbending of the phonon
dispersion of diamond introduces large (second-order) isotopic disorder
effects which are not present in the case of Si, Ge, and $\alpha$-Sn because
of their vanishing DOS at the frequency of the Raman phonon. Such overbending
of the phonon dispersion is also encountered in SiC and will be discussed in
Sec.~\ref{sec:compounds}.

We also observe a slight but systematic upwards shift of $x_{{\rm max},\Gamma}$
of the CPA values for Si, compared to those obtained by perturbation theory.
The points of the CPA calculation, which includes higher-order perturbation
terms, are fitted with Eq.~(\ref{gn-fit}) for $n=2,3$ that only contains the
second-order and third-order terms. As shown in Fig.~\ref{fig:gn}, the
higher-order terms (such as, e.g., $g_5$) emphasize even more the asymmetric
appearance of the self-energy, provided that the prefactors $A_i$ are all
positive. A fit with Eq.~(\ref{gn-fit}) for $n=2,3,4$ to the CPA values for the
disorder-induced broadening of Si can be found in
Ref.~[\onlinecite{Widulle01}].

\section{Extension for binary compounds}
\label{sec:compounds}

The total isotopic-disorder-induced self-energy for a phonon of a binary
compound~\cite{Tamura84} can be expressed as a sum of the self-energy
contributions of the individual sublattices $\kappa$

\begin{eqnarray}
\Pi_n(\omega,x)
&=&
\sum\limits_{\kappa} \frac{g_n^{\kappa}(x)}{2^n}\,\omega
\,|{\bf e}^{\kappa}(\omega)|^2\nonumber\\ & &
\times\left(\frac{1}{3N_c}\sum\limits_i \omega_i \,D(\omega,\omega_i)\,
|{\bf e}^{\kappa}(\omega_i)|^2 \right)^{n-1} \label{pi_n_bin}
\end{eqnarray}
where $|{\bf e}^{\kappa}|^2$ denotes the square of the eigenvector, averaged
over all ${\bf q}$'s corresponding to a given frequency. The second-order
self-energy can be described in a way similar to that used for monoatomic
semiconductors

\begin{mathletters}
\begin{eqnarray}
\label{delta2-binary}
\Delta_2(\omega,x)
&=& \sum\limits_{\kappa}
\frac{g_2^{\kappa}(x)}{4} \,\omega\,|{\bf e}^{\kappa}(\omega)|^2\nonumber\\
& & \times\left(
\frac{1}{3 N_c}\sum\limits_i
\omega_i\,|{\bf e}^{\kappa}(\omega_i)|^2\,
\frac{\omega-\omega_i}{(\omega-\omega_i)^2+\gamma^2} \right),\\
\label{gamma2-binary}
\Gamma_2(\omega,x) &=&\sum\limits_{\kappa}
\frac{g_2^{\kappa}(x)}{4} \,\omega\,|{\bf e}^{\kappa}(\omega)|^2\nonumber\\
& & \times \left(
\frac{1}{3 N_c}\sum\limits_i
\omega_i\,|{\bf e}^{\kappa}(\omega_i)|^2\,
\frac{\gamma}{(\omega-\omega_i)^2+\gamma^2} \right)
\end{eqnarray}

For $\gamma\rightarrow 0$, Eq.~(\ref{gamma2-binary}) simplifies to
\begin{eqnarray}
\label{gamma2-binary-harmonic}
-\Gamma_2(\omega,x) &=& \sum\limits_{\kappa}
\frac{g_2^{\kappa}(x)}{4} \,\omega\,|{\bf e}^{\kappa}(\omega)|^2\nonumber\\
& & \times\left( \frac{1}{3 N_c} \sum\limits_i
\omega_i\,|{\bf e}^{\kappa}(\omega_i)|^2\,\pi\delta(\omega -
\omega_i)\right)\nonumber\\ &=&\sum\limits_{\kappa} \frac{\pi}{12}\,
g_2^{\kappa}(x)\,\omega^2\,|{\bf
e}^{\kappa}(\omega)|^2\,\rho_1^{\kappa}(\omega), \end{eqnarray}
\end{mathletters}

where $\rho_1^{\kappa}(\omega)$ corresponds to the one-phonon partial DOS
projected on the sublattice $\kappa$. Note in
Eq.~(\ref{gamma2-binary-harmonic}) that for crystals with diamond structure
Eq.~(\ref{gamma_2-formula}) is recovered by setting $|{\bf
e}^{\kappa}(\omega)|^2=1/2$.

The general expression for $r_s$ can be quite complicated for binary
compounds when both sublattices contain isotopic disorder. However, if there
is only one isotopically disordered sublattice, the sum over $\kappa$ is
replaced by a single term. In this case, it is straightforward to derive
expressions similar to Eq.~(\ref{r3delta-final}) and (\ref{r3gamma-final}),

\begin{mathletters}
\begin{eqnarray}
\label{r3delta-finalbin}
r_{\Delta}
&\simeq &
~\,8\,\frac{m}{\Delta m}\,\frac{1}{\omega\,|{\bf e} (\omega)|^2}\,
\left[\frac{\Delta_2^2-\Gamma_2^2}{\Delta_2}\right]_{x=0.5}\\
\label{r3gamma-finalbin}
r_{\Gamma}
&\simeq &
16\,\frac{m}{\Delta m}\,\frac{1}{\omega\,|{\bf e}
(\omega)|^2}\,\left[\Delta_2\right]_{x=0.5}
\end{eqnarray}
\end{mathletters}

where ${\bf e}(\omega)$ is the eigenvector component at the disordered
sublattice for the phonon under consideration (e.g. a Raman phonon).
Thus, the isotopic effects of a disordered sublattice in a compound is
different from that for the corresponding monoatomic crystal.

We apply Eq.~(\ref{r3delta-finalbin}) and Eq.~(\ref{r3gamma-finalbin}) to the
Raman spectroscopic results on a variety of
${^{\rm{nat}}\rm{Si}}{^{12}\rm{C}_{1-x}}{^{13}\rm{C}_x}$ polytypes, recently
reported by Rohmfeld {\it et al}~[\onlinecite{Rohmfeld01}]. We have performed
a fit with Eq.~(\ref{gn-fit}) for $n=2,3$ to the linewidth of the transverse
optic (TO) modes of the 6H-SiC polytype measured in
Ref.~[\onlinecite{Rohmfeld01}] for $^{13}$C-concentrations ranging from
$x=0.15$ to $x=0.40$. These experimental results are displayed together with
our fits in Fig.~\ref{fig:SiCGamma}. The fit performed in
Ref.~[\onlinecite{Rohmfeld01}] was based on the asymmetric curve obtained by
CPA calculations for diamond.~\cite{Hass92} This curve was first fitted to the
data points of the TO(2/6) mode and further adjusted to the TO(0) and TO(6/6)
modes by multiplication with the constant scaling factor
$\omega^2\rho_1(\omega)$ assuming constant eigenvectors (see
Eq.~(\ref{gamma2-binary-harmonic}). Instead, we have considered each TO mode
separately and performed fits with Eq.~(\ref{gn-fit}) for $n=2,3$ in the same
manner as for the elemental semiconductors in Sec.~\ref{sec:comparison}. We
used, however, parameters appropriate to SiC, not to diamond. In this way, we
rigorously conclude that the behavior of $\Gamma_{\rm dis}$ versus $x$ is
asymmetric. This fact cannot be derived from the data in Fig.~3 of
Ref.~[\onlinecite{Rohmfeld01}] which were obtained only for $0.15<x<0.40$. The
latter can be fitted equally with either a symmetric or an asymmetric curve.

Using Eq.~(\ref{gamma2-binary-harmonic}) for a single disordered sublattice,
together with the assumption of a constant eigenvector $|{\bf e}_{\rm TO}^{\rm
C}({\bf q})| = 0.84$,\cite{note2} we have calculated the imaginary part
$\Gamma_2(\omega)$ for 6H-SiC and 3C-SiC using the DOS of Hofmann {\it et
al}.~[\onlinecite{Hofmann94}]. The corresponding real part,
$\Delta_2(\omega)$, is obtained by the Kramers-Kronig relations. The frequency
dependence of $\Delta_2$ and $\Gamma_2$ for 6H-SiC, as well as $\Gamma_2$ for
zincblende SiC (3C-SiC) is shown in Fig.~\ref{fig:SiCBCM} for the region of
the TO modes. The three parallel lines cut the $x$-axis at the frequencies of
the TO(0), TO(2/6) and TO(6/6) phonons, and the crossing with $\Delta_2(6H)$
gives the (approximate) disorder-induced shift of the corresponding Raman peak.

Raman experimental results extracted from Fig.~\ref{fig:SiCGamma} are
compared in Table~\ref{tab:comparisonSiC} with those found by the
perturbation scheme of Eq.~(\ref{r3delta-finalbin}) and
(\ref{r3gamma-finalbin}). The fit to the TO(0) mode is symmetric, but the
large scattering of data points still allows reasonable asymmetric fits with
similar curvature than for the other two modes. The two numbers given for the
ratio $r_\Delta$ of the TO(0) mode represent maximum and minimum values and
correspond to 6H-SiC and 3C-SiC, respectively. For the cubic polytype the DOS
vanishes at the Raman frequencies in the harmonic approximation. Therefore, we
performed a convolution of the imaginary part with the anharmonic broadening
$\Gamma_{\rm anh} = 1.4$~cm$^{-1}$~[\onlinecite{Ulrich99}] in order to estimate
the finite value. Note that the phonon dispersion of the hexagonal polytypes,
calculated in Ref.~[\onlinecite{Hofmann94}], exhibits an overbending of the TO
branches, which mainly occurs around the K-point of the Brillouin zone. The
relatively large value of $\Gamma_2$ observed for the TO(0) mode compared to
those of Si, Ge and $\alpha$-Sn (see Table~\ref{tab:comparison}) provides
evidence of a non-vanishing DOS caused by an overbending of the phonon
dispersion. The experimental results from Ref.~[\onlinecite{Rohmfeld01}]
support therefore the DOS calculated with the bond charge model by Hofmann {\it
et al}.~[\onlinecite{Hofmann94}].

The striking feature in Table~\ref{tab:comparisonSiC}, which is added to the
general behavior of the Raman phonons of elemental crystals, is the negative
values of $r_\Delta$ obtained for the TO(2/6) and TO(6/6) phonons.  This
change of sign implies a reversed asymmetry, i.e. the maximum is now shifted to
lower concentrations of the heavier isotope ($x<0.5$). Since in
Ref.~[\onlinecite{Rohmfeld01}] the phonon frequency was used for the
determination of the $^{13}$C-concentration, no experimental information
can be exctracted about the disorder-induced shift. Within our interpretation,
however, a negative $r_\Delta$ has been already observed indirectly for a
two-phonon combination in isotopically tailored diamond by
cathodoluminescence.\cite{Ruf98} The fact that some absolute values of
$r_{\Delta,\Gamma}$ are much larger than unity is caused by a strong difference
in magnitude between $\Delta_2$ and $\Gamma_2$ or by a vanishing $\Delta_2$
(see Eq.~(\ref{r3delta-finalbin}) and Eq.~(\ref{r3gamma-finalbin})). The latter
occurs at frequencies near the maximum of the phonon DOS where $\Delta(\omega)$
changes sign.

\section{Conclusions}
General analytic expressions are given which allow the calculation of the real
and imaginary parts of the self-energy of phonons due to isotopic disorder in
elemental crystals (e.g. diamond-type semiconductors) to various orders of
perturbation theory. The results are used to analyze the asymmetric
behavior of these self-energies versus the isotopic concentration in crystals
containing two different isotopes. The asymmetry is found to originate
primarily from a combination of second-order and third-order contributions to
the phonon self-energies and can be determined from the second-order
perturbation terms without the need of additional parameters. These results are
extended to binary crystals (e.g. zincblende or wurtzite-type semiconductors)
and used to interpret quantitatively recent experimental data for SiC versus
the isotopic composition of the carbon component. The possible asymmetric
behavior of the self-energy in this material, which contrary to the
statement in Ref.~[\onlinecite{Rohmfeld01}] does not follow from the available
data for $0.15<x<0.40$ only, is clarified on the basis of our analytical
expressions.

\acknowledgments
We would like to thank W.~Kress for a careful reading of the manuscript. J.~S.
acknowledges support from the Max-Planck-Gesellschaft and the Ministerio    
de Educaci\'on y Cultura (Spain) through the Plan Nacional de Formaci\'on del
Personal Investigador.

\begin{figure}
\begin{minipage}{13cm}
\epsfxsize=13cm
\centerline{\epsffile{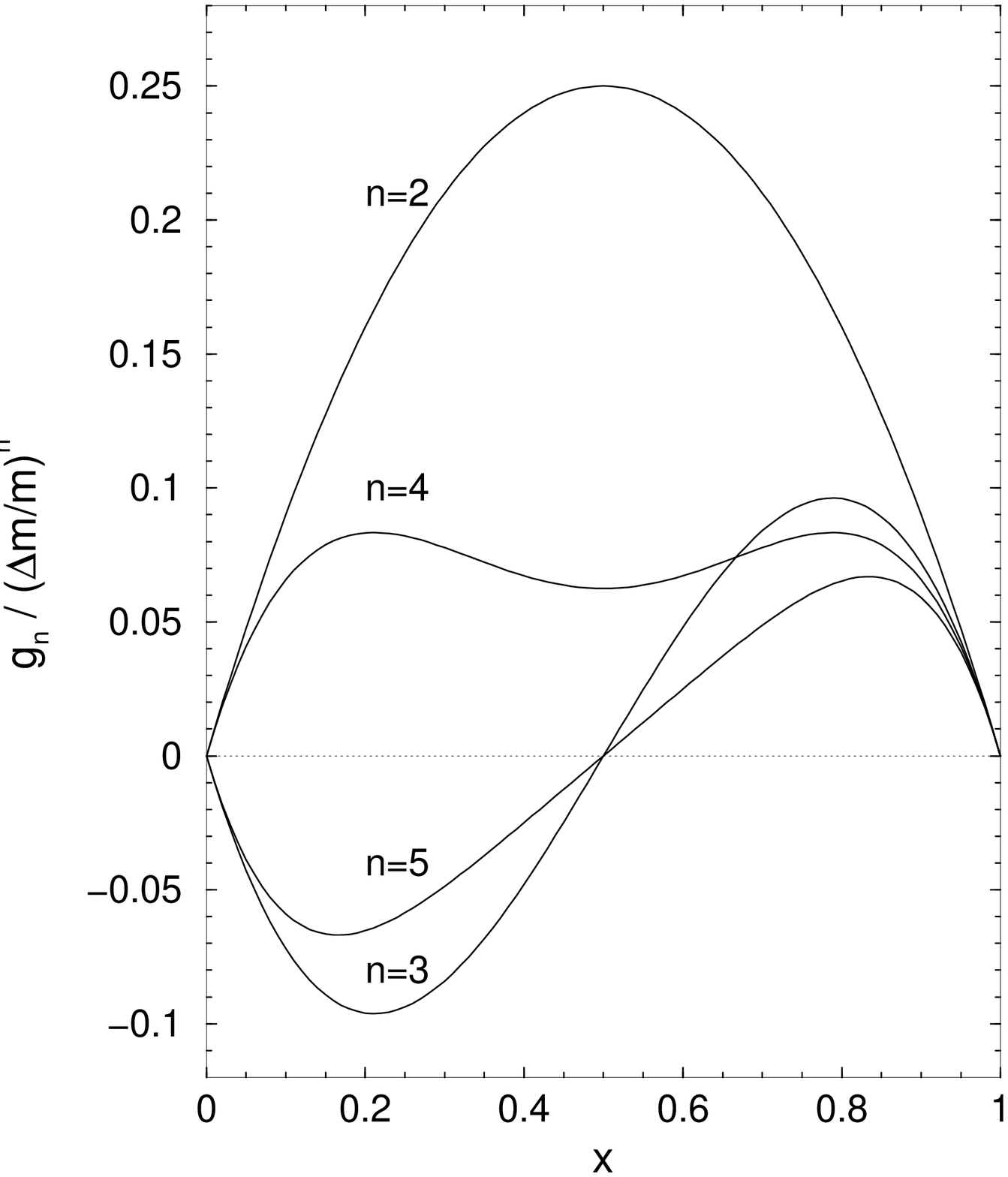}}
FIG.~\ref{fig:gn}
\end{minipage}
\end{figure}

\begin{figure}
\begin{minipage}{13cm}
\epsfxsize=13cm
\centerline{\epsffile{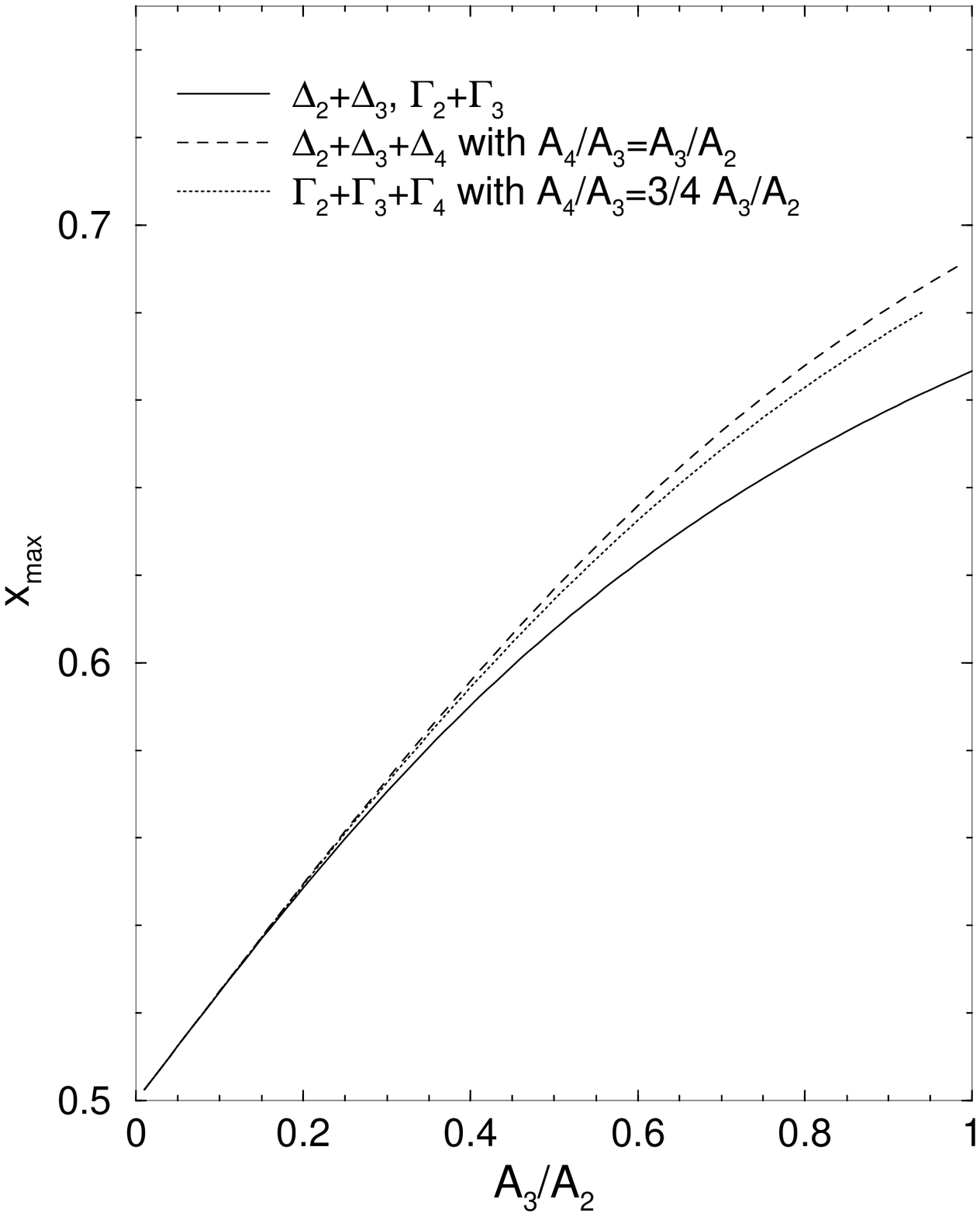}}
FIG.~\ref{fig:xmax}
\end{minipage}
\end{figure}

\begin{figure}
\begin{minipage}{13cm}
\epsfxsize=13cm
\centerline{\epsffile{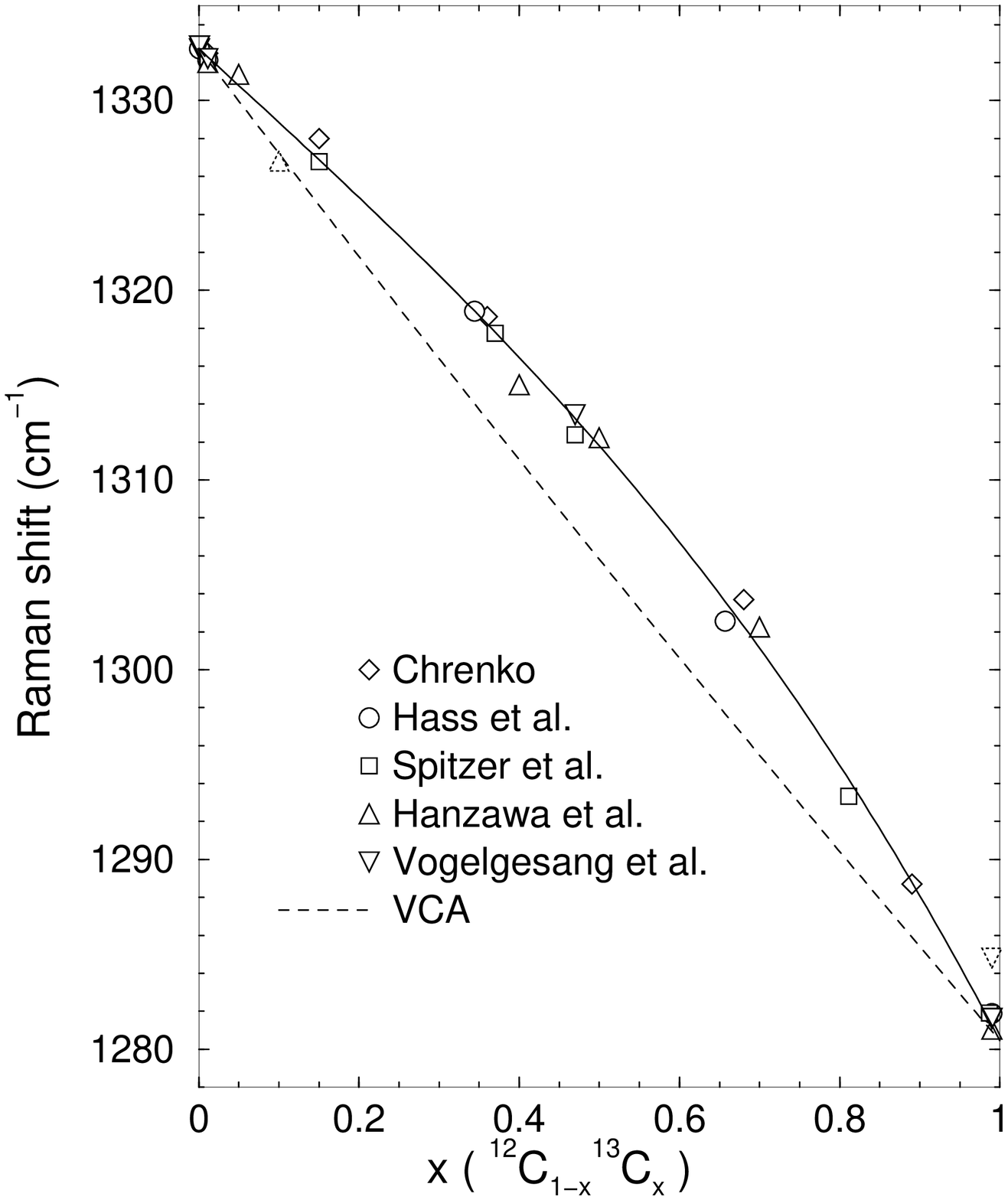}}
FIG.~\ref{fig:DiamondShift}
\end{minipage}
\end{figure}

\begin{figure}
\begin{minipage}{13cm}
\epsfxsize=13cm
\centerline{\epsffile{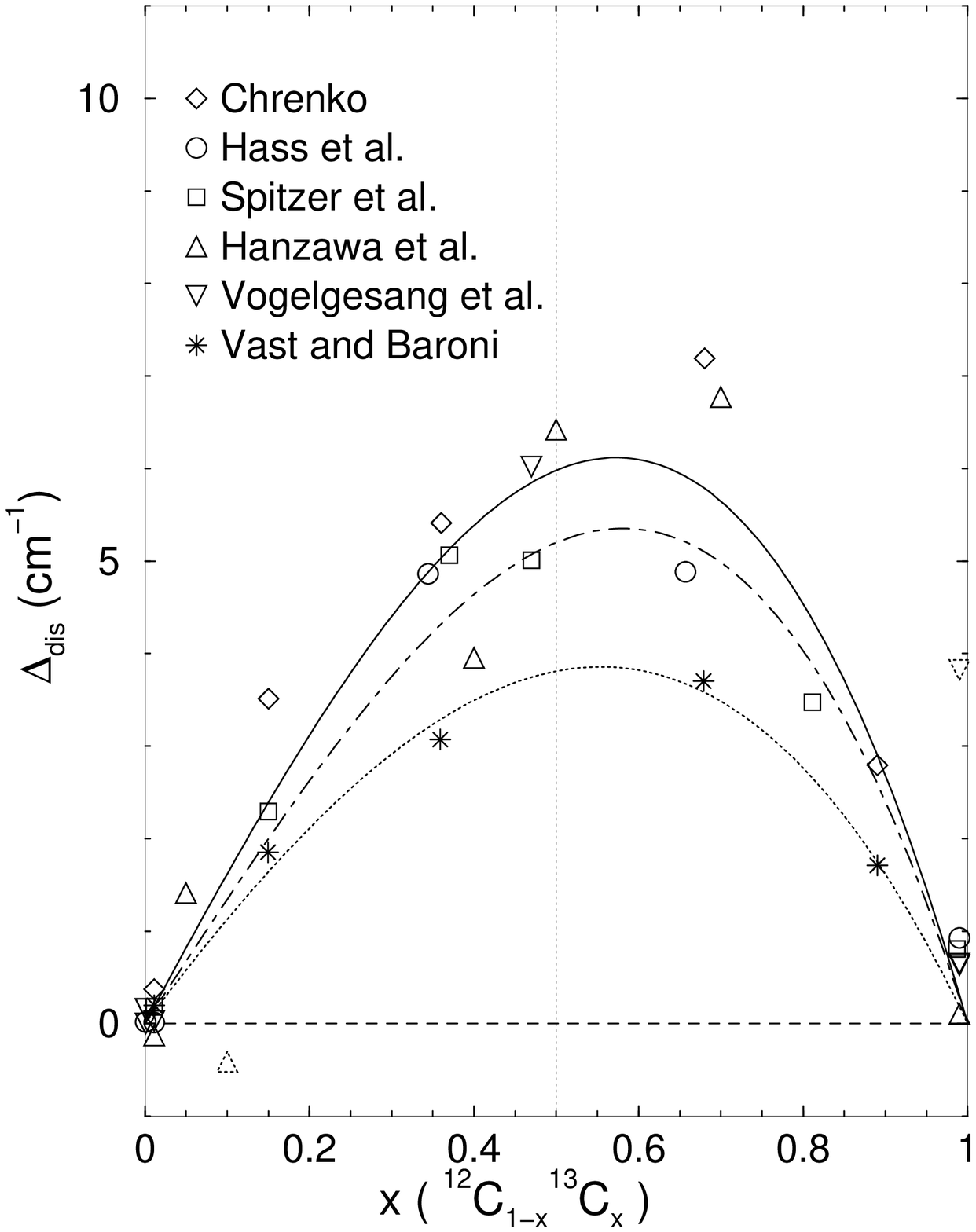}}
FIG.~\ref{fig:DiamondDelta}
\end{minipage}
\end{figure}

\begin{figure}
\begin{minipage}{13cm}
\epsfxsize=13cm
\centerline{\epsffile{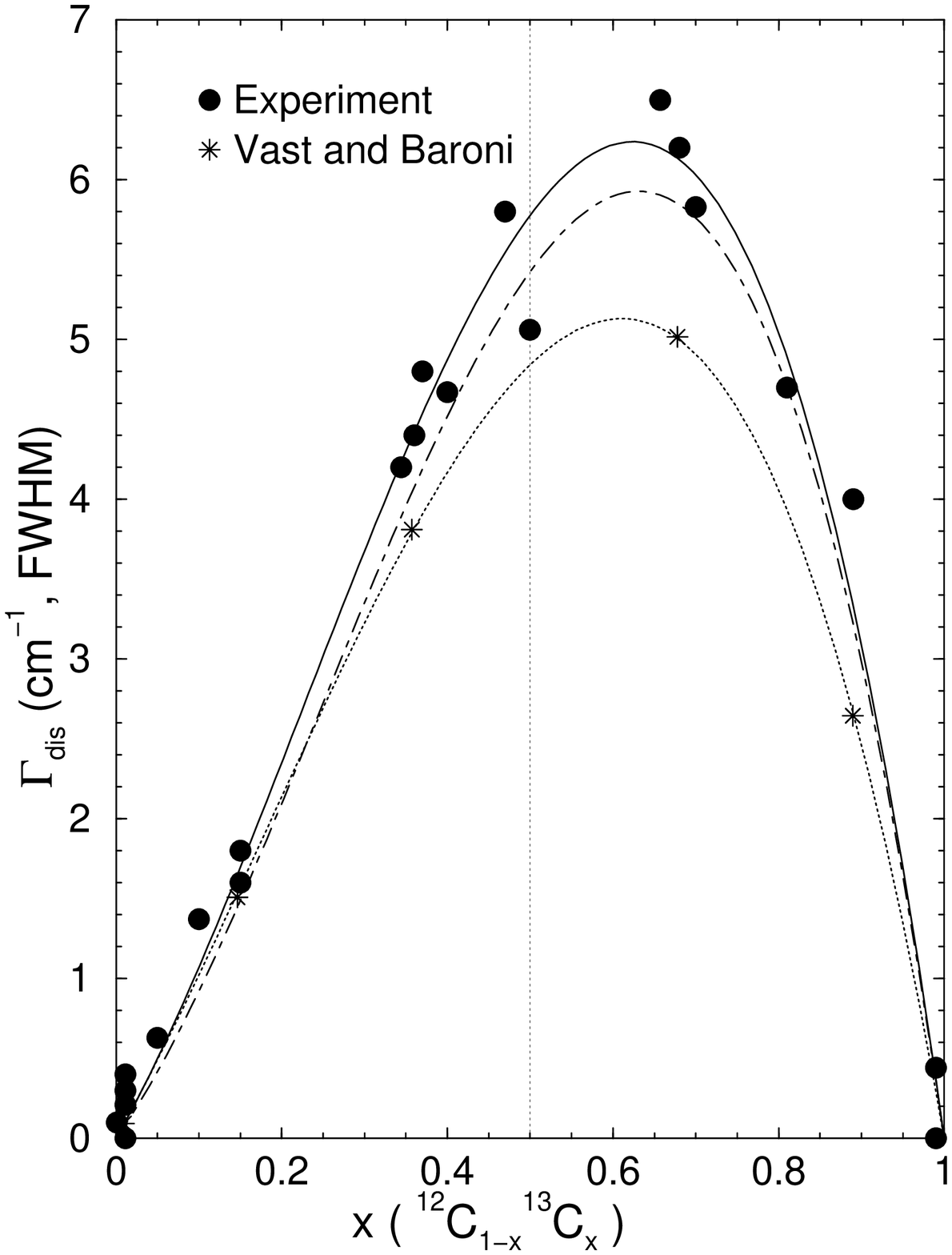}}
FIG.~\ref{fig:DiamondGamma}
\end{minipage}
\end{figure}

\begin{figure}
\begin{minipage}{13cm}
\epsfxsize=13cm
\centerline{\epsffile{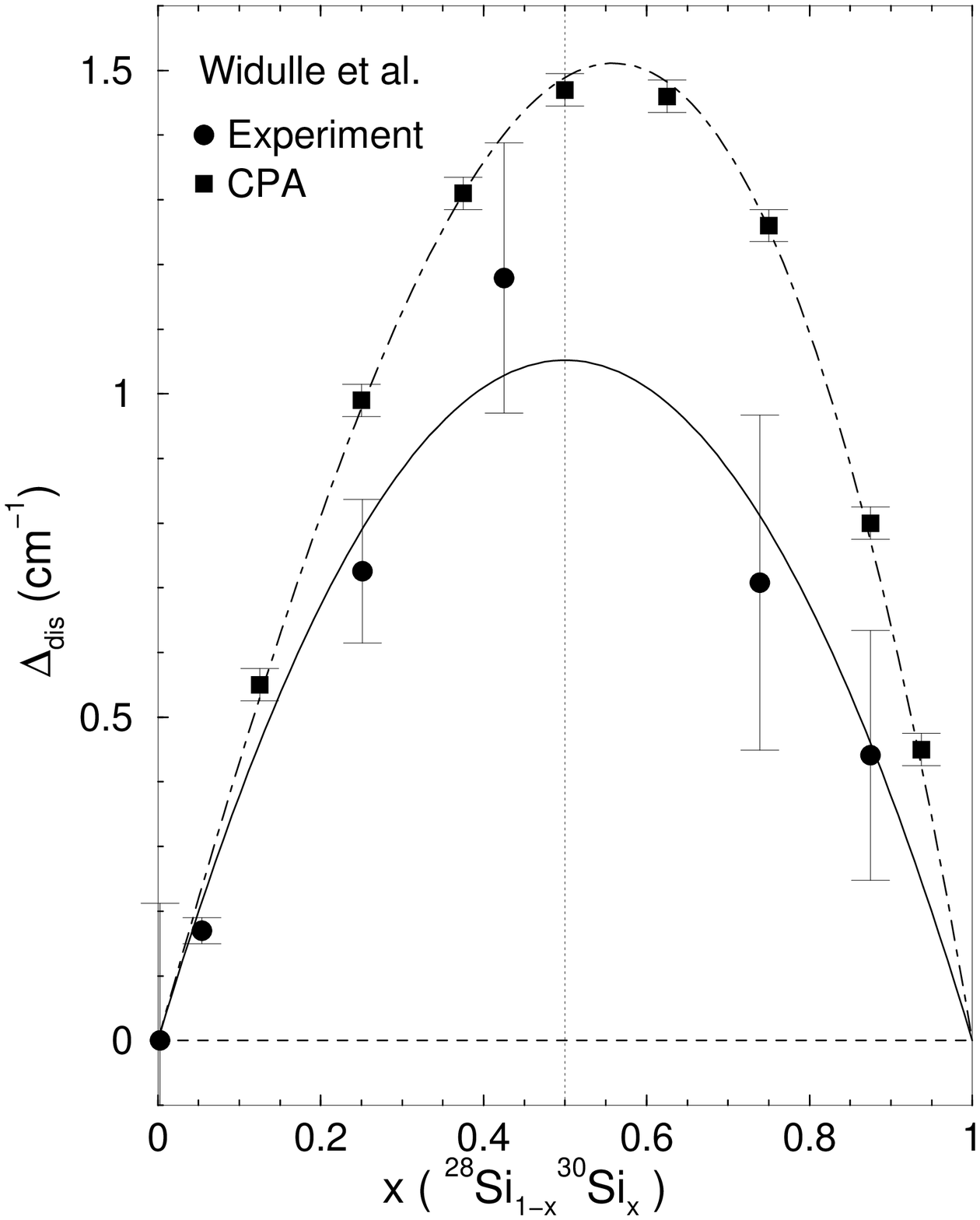}}
FIG.~\ref{fig:Si_Delta}
\end{minipage}
\end{figure}

\begin{figure}
\begin{minipage}{13cm}
\epsfxsize=13cm
\centerline{\epsffile{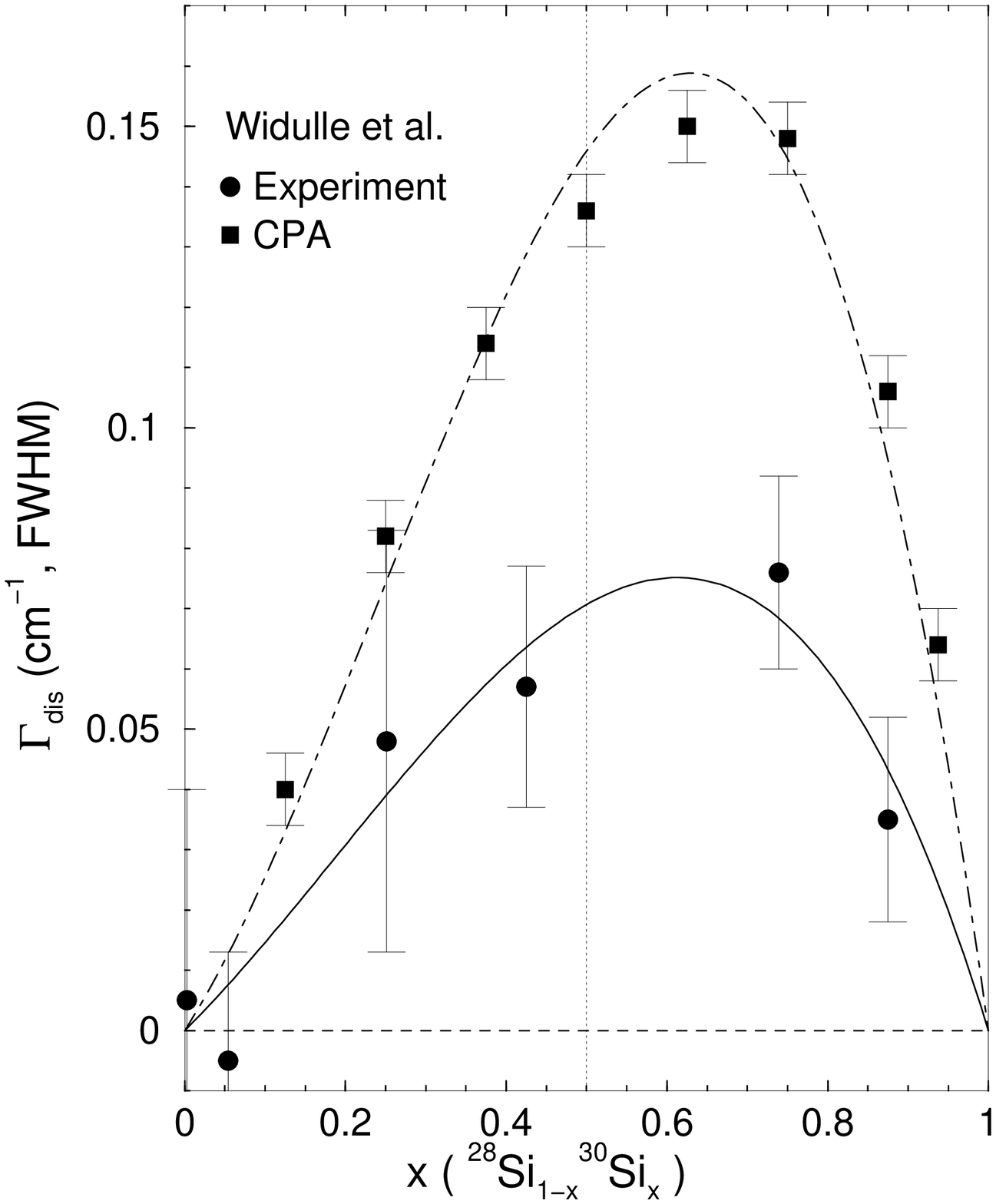}}
FIG.~\ref{fig:Si_Gamma}
\end{minipage}
\end{figure}

\begin{figure}
\begin{minipage}{13cm}
\epsfxsize=13cm
\centerline{\epsffile{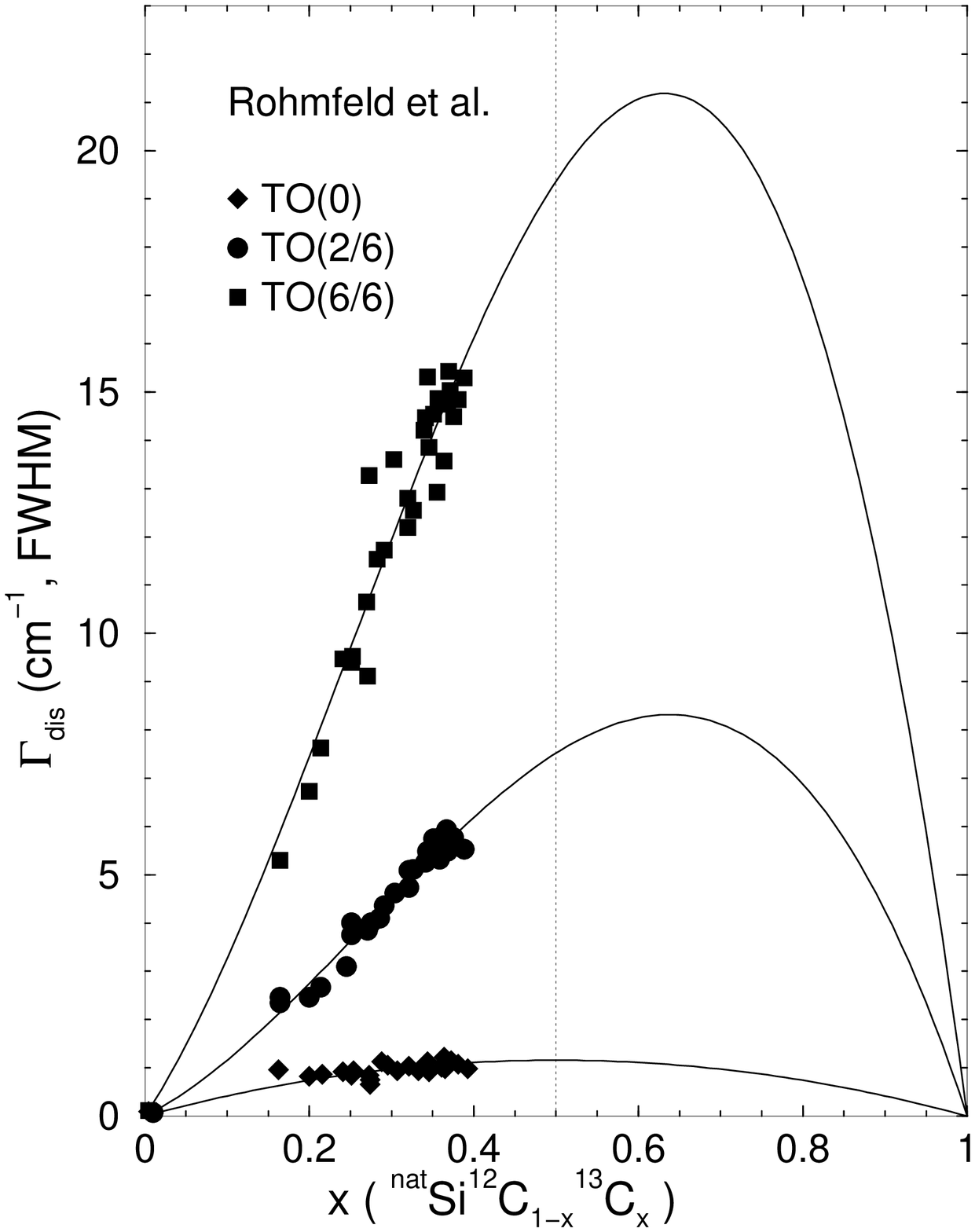}}
FIG.~\ref{fig:SiCGamma}
\end{minipage}
\end{figure}

\begin{figure}
\begin{minipage}{13cm}
\epsfxsize=13cm
\centerline{\epsffile{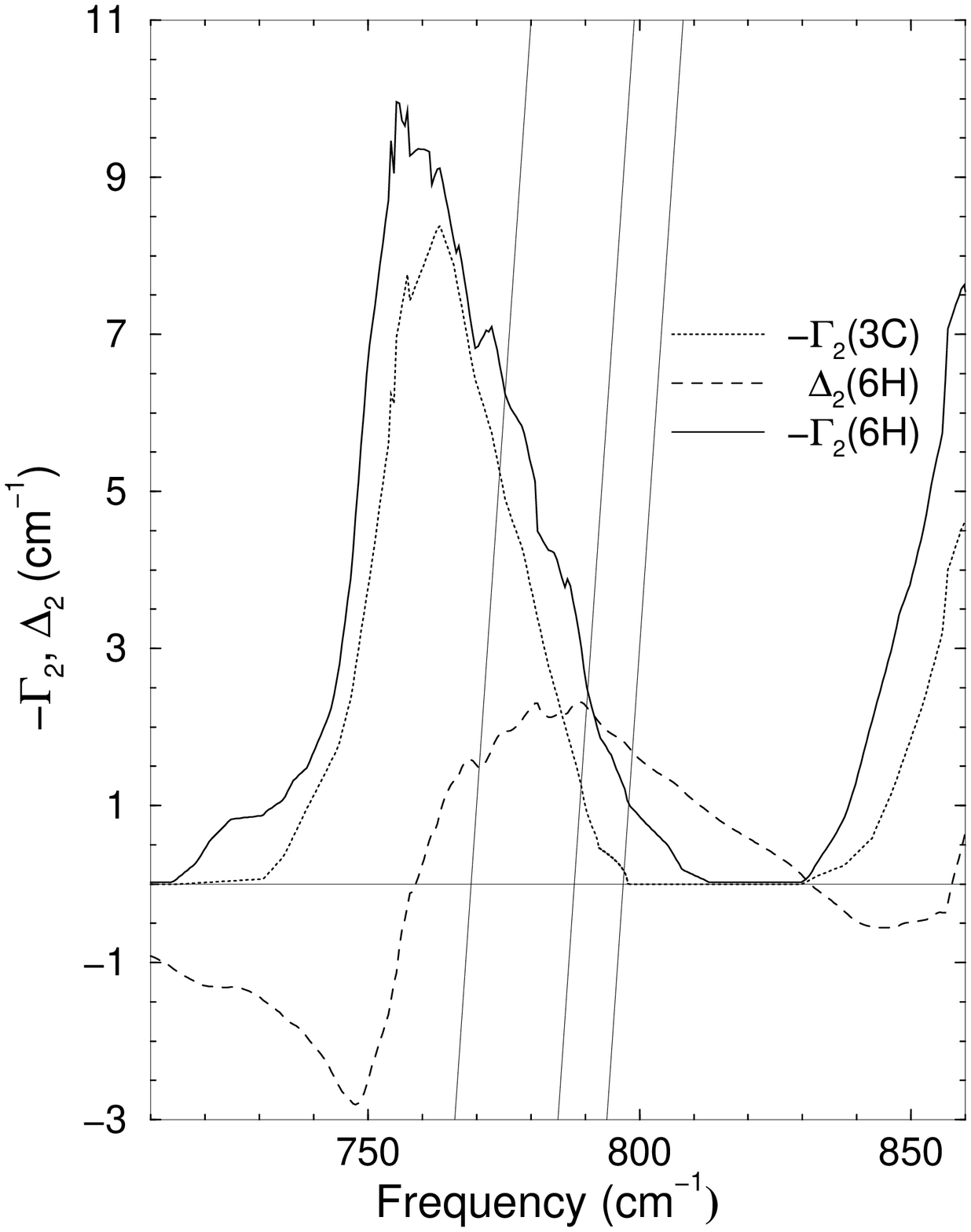}}
FIG.~\ref{fig:SiCBCM}
\end{minipage}
\end{figure}

\newpage

\begin{figure}
\caption[]{Mass-fluctuations $g_n(x)$ for various $n$-th moments
(Eqs.~(\ref{g2345})) of an elemental, i.e. monoatomic, crystal (or each
sublattice of a compound) containing a binary isotope system. $x$
denotes the concentration of the heavier mass isotope.}
\label{fig:gn}
\end{figure}

\begin{figure}
\caption[]{Dependence of the value of $x_{\rm max}$ at which the maxium of
$\Delta$ and $\Gamma$ is found, including second-order and third-order
perturbation terms (solid line) and also fourth-order terms (dashed and dotted
lines).}
\label{fig:xmax}
\end{figure}

\begin{figure}
\caption[]{Raman shift of isotopically disordered diamond. The open symbols
represent experimental
values~[\onlinecite{Chrenko88,Hass91,Hass92,Spitzer93,Hanzawa96,Vogelgesang96}]
The dashed line indicates the harmonic scaling of the phonon frequency within
the VCA ($\omega\propto\overline{m}^{-1/2}$). The solid line corresponds to a
fit with Eq.~(\ref{gn-fit}) for $n=2,3$ to all experimental data, added to the
VCA-scaling. The two experimental data points at $x=0.1$ and $x=1$, represented
as dotted symbols, were excluded from this fit.}
\label{fig:DiamondShift}
\end{figure}

\begin{figure}
\caption[]{Disorder-induced shift of the Raman phonon of diamond as a function
of the $^{13}\rm{C}$-concentration. The open symbols are
Raman experimental
data~[\onlinecite{Chrenko88,Hass91,Hass92,Spitzer93,Hanzawa96,Vogelgesang96}]
whereas the asterisks correspond to {\it ab initio}
calculations~[\onlinecite{Vast00a,Vast00b}]. The solid line is a fit with
Eq.~(\ref{gn-fit}) for $n=2,3$ to all experimental data except for the two
points at $x=0.1$ and $x=1$ indicated as dotted symbols. The dotted and
dot-dashed lines represent the fits to theoretical values obtained from {\it
ab initio} and CPA calculations, respectively.}
\label{fig:DiamondDelta}
\end{figure}

\begin{figure}
\caption[]{Disorder-induced broadening of the Raman phonon of diamond as
a function of the $^{13}\rm{C}$-concentration. The filled circles have been
obtained from the Raman data of
Refs.~[\onlinecite{Chrenko88,Hass91,Hass92,Spitzer93,Hanzawa96}] by
taking into account the corresponding instrumental resolutions and subtracting
the anharmonic broadening, $\Gamma_{\rm anh} \approx 2~\rm{cm}^{-1}$ (FWHM).
The solid line is a fit with Eq.~(\ref{gn-fit}) for $n=2,3$ to these points.
The dotted and dot-dashed lines are the corresponding fits to the values
obtained from {\it ab initio}~[\onlinecite{Vast00a,Vast00b}] and
CPA~[\onlinecite{Hass91,Hass92}] calculations, respectively.}
\label{fig:DiamondGamma}
\end{figure}

\begin{figure}
\caption[]{Disorder-induced shift of the Raman phonon of Si as a function of
the $^{30}\rm{Si}$-concentration~[\onlinecite{Widulle01}]. The solid line is a
fit with Eq.~(\ref{gn-fit}) for $n=2,3$ to the experimental data. The
dot-dashed line represents the corresponding fit to the values obtained from
CPA calculations.}
\label{fig:Si_Delta}
\end{figure}

\begin{figure}
\caption[]{Disorder-induced broadening of the Raman phonon of Si as a function
of the $^{30}\rm{Si}$-concentration~[\onlinecite{Widulle01}]. The solid line
is a fit with Eq.~(\ref{gn-fit}) for $n=2,3$ to the experimental data. The
dot-dashed line represents the corresponding fit to the values obtained from
CPA calculations.}
\label{fig:Si_Gamma}
\end{figure}

\begin{figure}
\caption[]{Disorder-induced broadening of the TO Raman modes of the 6H-SiC
polytype versus the $^{13}{\rm C}$-concentration of the carbon sublattice.
The data are taken from Ref.~[\onlinecite{Rohmfeld01}]. The solid lines
represent fits with Eq.~(\ref{gn-fit}) for $n=2,3$ to the data points that
correspond to the TO(0), TO(2/6), and TO(6/6) phonon modes.}
\label{fig:SiCGamma}
\end{figure}

\begin{figure}
\caption[]{Disorder-induced shift $\Delta_2(\omega)$ and broadening
$\Gamma_2(\omega)$ for 3C-SiC and 6H-SiC. We obtain $\Gamma_2(\omega)$ from the
DOS that has been calculated by Hofmann {\it et al.}~[\onlinecite{Hofmann94}]
using the bond-charge model. $\Delta_2(\omega)$ is determined by a
Kramers-Kronig transformation.}
\label{fig:SiCBCM}
\end{figure}

\begin{table}[t]
\caption[]{Characterization of the asymmetry of the phonon self-energy with
respect to $x=0.5$ through the ratio $r_s=A_{3,s}/A_{2,s}$ and the
concentration $x_{{\rm max},s}$ with $s=\Delta,\Gamma$. These quantities are
displayed for each elemental crystal that is realized as a binary isotopic
system $m_{1-x}M_x$, with separate rows for the corresponding real and
imaginary part. The Raman data are compared with theoretical results that are
derived from perturbation theory (PTh, this work), the coherent potential
approximation (CPA), and {\it ab initio} calculations. The values of the
self-energies ($\Delta_2,\Gamma_2$) are for $x=0.5$.}
\label{tab:comparison}

\begin{tabular}{lddddddddd}

 & \multicolumn{3}{c}{Raman experiments} & \multicolumn{2}{c}{PTh} &
\multicolumn{2}{c}{CPA} & \multicolumn{2}{c}{\it ab initio$^e$} \\\hline

 & \multicolumn{1}{r}{$\Delta_2~(\rm{cm}^{-1})$} & $r_{\Delta}$ &
$x_{\rm{max},\Delta}$  & $r_{\Delta}$ & $x_{\rm{max},\Delta}$  & $r_{\Delta}$ &
$x_{\rm{max},\Delta}$  & $r_{\Delta}$ & $x_{\rm{max},\Delta}$\\
\raisebox{1.5ex}[-1.5ex]{${\rm{m}_{1-x}}{\rm{M}_x}$} &
\multicolumn{1}{r}{$-\Gamma_2~(\rm{cm}^{-1})$} & $r_{\Gamma}$ &
$x_{\rm{max},\Gamma}$ & $r_{\Gamma}$ & $x_{\rm{max},\Gamma}$  & $r_{\Gamma}$ &
$x_{\rm{max},\Gamma}$  & $r_{\Gamma}$ & $x_{\rm{max},\Gamma}$\\\hline
\\
 & $\approx$ 6$^a$ & 0.31 & 0.57 & 0.33 & 0.58 & 0.35 & 0.58 & 0.22 &
0.55\\ \raisebox{1.5ex}[-1.5ex]{${^{12} \rm{C}_{1-x}}{^{13} \rm{C}_x}$}
 & $\approx$ 3$^a$ & 0.61 & 0.62 & 0.88 & 0.66 & 0.66 & 0.63 &
0.53 & 0.61 \\ \\
 & 1.18$^b$ & $\approx$ 0 & 0.50 & 0.26 & 0.56 & 0.25 & 0.56 & &\\
\raisebox{1.5ex}[-1.5ex]{${^{28}\rm{Si}_{1-x}}{^{30} \rm{Si}_x}$}
 & 0.03$^b$ & 0.53 & 0.61 & 0.52 & 0.61 & 0.64 & 0.63 & &\\

 &1.06$^c$ & & & 0.33 & 0.58 & & & 0.30 & 0.57\\
\raisebox{1.5ex}[-1.5ex]{${^{70} \rm{Ge}_{1-x}}{^{76} \rm{Ge}_x}$}
 &0.03$^c$ & & & 0.66 & 0.63 & & & 0.74 & 0.64\\

 &1.8$^d$ & & & 0.67 & 0.63 & & & &\\
\raisebox{1.5ex}[-1.5ex]{${^{112}\rm{Sn}_{1-x}}{^{124}\rm{Sn}_x}$}
 &0.02$^d$ & & & $\approx$ 1 & $\approx$ 0.67 & & & &\\

 &0.7$^d$ & & & 0.41 & 0.59 & & & &\\
\raisebox{1.5ex}[-1.5ex]{${^{116}\rm{Sn}_{1-x}}{^{124}\rm{Sn}_x}$}
 &0.02$^d$ & &  & 0.82 & 0.65 & & & &

\end{tabular}
\tablenotetext{$^a$Values taken as average from
Refs.~[\onlinecite{Chrenko88,Hass91,Hass92,Spitzer93,Hanzawa96,Vogelgesang96}].} \tablenotetext{$^b$Ref.~[\onlinecite{Widulle01}].}
\tablenotetext{$^c$Ref.~[\onlinecite{Zhang98}].}
\tablenotetext{$^d$Ref.~[\onlinecite{Wang97}].}
\tablenotetext{$^e$Refs.~[\onlinecite{Vast00a,Vast00b}].}
\end{table}

\begin{table}[t]
\caption[]{Asymmetry of the phonon self-energies of isotopically disordered
SiC, characterized by the ratio $r_s=A_{3,s}/A_{2,s}$ and the concentration
$x_{{\rm max},s}$. Experimental results from Raman spectroscopy of the
TO-phonons of 6H-SiC~[\onlinecite{Rohmfeld01}] are compared with calculations
using perturbation theory (PTh, this work). The values of the self-energies
($\Delta_2,\Gamma_2$) are for $x=0.5$.}
\label{tab:comparisonSiC}

\begin{tabular}{ldddddd}

 & \multicolumn{3}{c}{Raman experiment} & \multicolumn{3}{c}{PTh$^*$}\\\hline

 & \multicolumn{1}{r}{$\Delta_2~(\rm{cm}^{-1})$} & $r_{\Delta}$ &
$x_{\rm{max},\Delta}$ & \multicolumn{1}{r}{$\Delta_2~(\rm{cm}^{-1})$} &
$r_{\Delta}$ & $x_{\rm{max},\Delta}$\\ \raisebox{1.5ex}[-1.5ex]{
${^{\rm{nat}}\rm{Si}}{^{12}\rm{C}_{1-x}}{^{13}\rm{C}_x}$} &
\multicolumn{1}{r}{$-\Gamma_2~(\rm{cm}^{-1})$} & $r_{\Gamma}$ &
$x_{\rm{max},\Gamma}$ & \multicolumn{1}{r}{$-\Gamma_2~(\rm{cm}^{-1})$} &
$r_{\Gamma}$ & $x_{\rm{max},\Gamma}$\\\hline \\
 & -- & & & 1.7 & 0.28-0.21 & 0.57\\
\raisebox{1.5ex}[-1.5ex]{TO(0)} & $\approx$0.55 & $\approx 0$ & $\approx$0.5 &
0.12-0.90 & 0.58 & 0.62\\

 & -- & & & 2.2 & -0.75 & 0.36\\
\raisebox{1.5ex}[-1.5ex]{TO(2/6)} & $\approx$3.8 & 0.71 & 0.64 & 3.8 & 0.76 &
0.64 \\

 & -- & & & 1.5 & -6.0 & \\
\raisebox{1.5ex}[-1.5ex]{TO(6/6)} & $\approx$9.7 & 0.66 & 0.63 & 7.3 & 0.53 &
0.61

\end{tabular}
\tablenotetext{$^*$ Note that for negative values the asymmetry is flipped with
respect to $x=0.5$, i.e. the maximum of the self-energy occurs at $x<0.5$.}
\end{table}


\begin{references}
\bibitem{note1}
Note that $-\Gamma/2$ is positive whereas $\Delta$ can be either positive or
negative.
\bibitem{note2}
For the TO modes of SiC, the variation of the eigenvectors $|e(\mathbf{q})|$
with $\mathbf{q}$ is small.\cite{Hofmann94,Karch94,Widulle99}
\end{references}
\end{document}